\documentclass[twocolumn,aps,showpacs,floatfix,prc]{revtex4-1}
\usepackage[dvips]{epsfig}

\begin{document}

\title {Isospin asymmetric nuclear matter and properties of axisymmetric neutron stars}

\author{Partha Roy Chowdhury$^1$}
\email{royc.partha@gmail.com} 
\author{Abhijit Bhattacharyya$^1$}
\email{abphy@caluniv.ac.in}
\author{D. N. Basu$^2$}
\email{dnb@veccal.ernet.in}

\affiliation{$^1$Department of Physics, University of Calcutta, 92 A.P.C. Road, Kolkata-700 009, India\\
            $^2$Variable  Energy  Cyclotron  Centre, 1/AF Bidhan Nagar, Kolkata-700 064, India }

\date{\today }

\begin{abstract}

    Pure hadronic compact stars, above a limiting value ($\approx$1.6 M$_\odot$) of their gravitational masses, to which predictions of most of other equations of state (EoSs) are restricted, can be reached from the equation of state (EoS) obtained using DDM3Y effective interaction. This effective interaction is found to be quite successful in providing unified description of elastic and inelastic scattering, various radioactivities and nuclear matter properties. We present a systematic study of the properties of pure hadronic compact stars. The $\beta$-equilibrated neutron star matter using this EoS with a thin crust is able to describe highly-massive compact stars, such as PSR B1516+02B with a mass M=1.94$^{+0.17}_{-0.19}$ M$_\odot$ and PSR J0751+1807 with a mass M=2.1$\pm$0.2  M$_\odot$ to a 1$\sigma$ confidence level.

\vskip 0.2cm
\noindent

\end{abstract}


\maketitle

\noindent

    The theoretical study of the nuclear equation of state is a field of research which ties together different areas of physics. Nuclear EoS is of great interest as its features control the stability of neutron star (NS), the evolution of the universe, supernova explosion, nucleosynthesis as well as central collisions of heavy nuclei. Extensive studies in the past two decades of nuclear matter created at subnormal or supernormal density in heavy ion collisions have resulted in experimental constraints on the nuclear EoS of symmetric matter. Recent astrophysical observations of massive neutron stars and heavy-ion data are confronted with our present understanding of the EoS of dense hadronic matter. We argue that the data from massive neutron stars and pulsars provide an important cross-check between high-density astrophysics and heavy-ion physics. The density dependence of nuclear symmetry energy (NSE) obtained by using nuclear EoS plays an important role for modelling the structure of the neutron stars (NSs) and the dynamics of supernova explosions since a series of observables (e.g. slope L of NSE, the value of NSE at nuclear density etc.) can be determined from the knowledge of symmetry energy. The stiffness of the high-density matter controls the maximum mass of compact stars. New measurements of the properties of pulsars point towards large masses and correspondingly to a rather stiff EoS \cite{La07} characterized by symmetric nuclear matter (SNM) incompressibility 250-270 MeV or more. In a recent analysis of x-ray burster EXO 0748-676 (M$=2.10\pm0.28$ M$_{\odot}$) it is even claimed that soft nuclear EoS are ruled out \cite{Oz06}. In addition, it is argued in Ref.\cite{Oz06} that condensates and deconfined quarks \cite{uma97} may not exist in the cores of NSs. Recently, new observations for the mass and the radius of compact stars have been obtained which provide stringent constraints on the EoS of strongly interacting matter at high densities \cite{Kl06,He00}. 

   We investigate the impacts of the compression modulus and symmetry energy of nuclear matter on the maximum mass of NSs in view of the recent constraints from the isospin diffusion in heavy-ion collisions at intermediate energies \cite{Ch05,Li08}. In the present work, the density dependent M3Y effective interaction (DDM3Y) \cite{Be77} which provides a unified description of the elastic and the inelastic scattering \cite{Gu05,Gu06}, various radioactivites \cite{BCS05,CSB06,prc07,scb07,Bas02} and nuclear matter properties \cite{BCS08,CBS09,BCS09}, is employed to obtain EoS of the $\beta$ equilibrated NS matter. This EoS is used to carry out a systematic study of the static as well as rotating NSs in view of the recent observations of the massive compact stars. 
\noindent
\begin{figure*}[]
\centerline{\includegraphics[width=3.38in,height=3.38in]{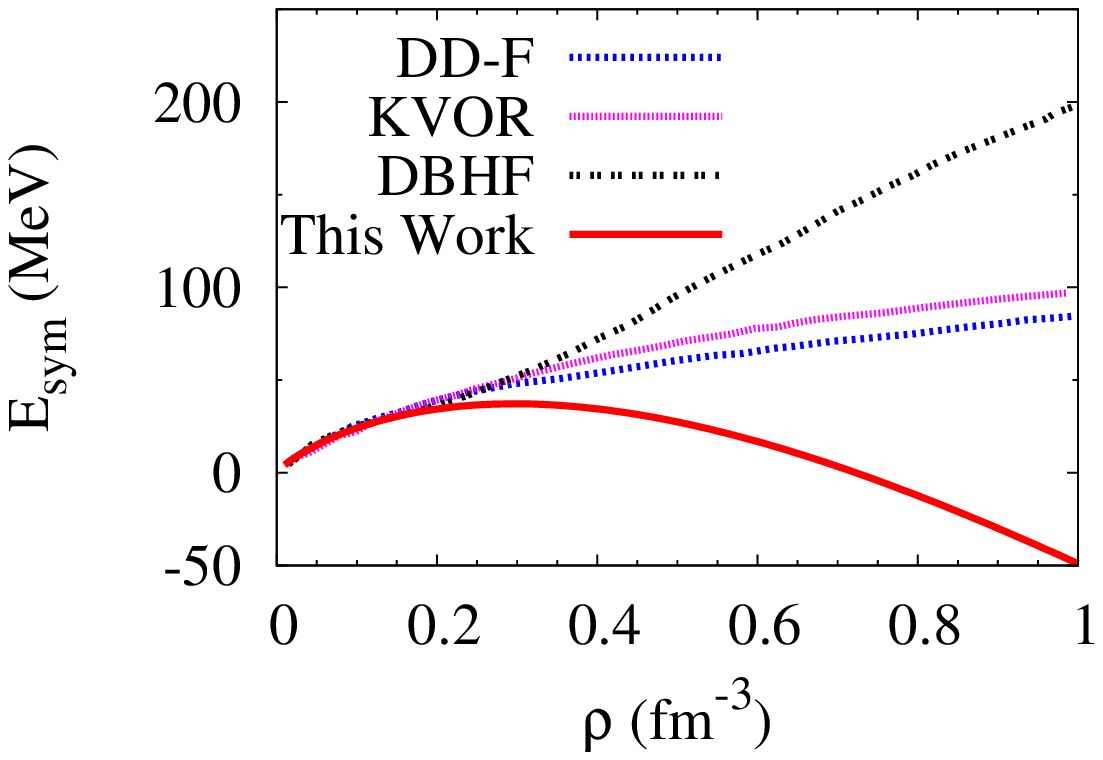}
\includegraphics[width=3.38in,height=3.38in]{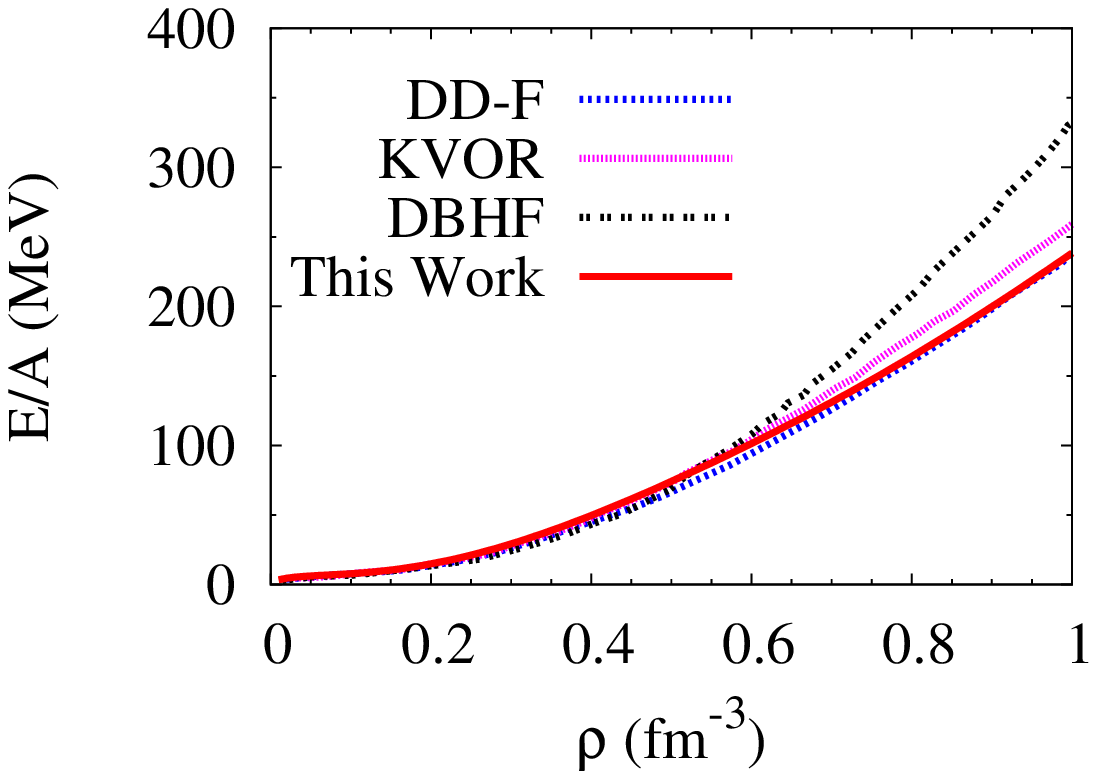}}
\caption{(Color online) (a) (Left) Density dependent behavior of symmetry energy (E$_{sym}$ vs. $\rho$) determined by different approaches.
(b) (Right) The EoS determined by different models (E/A vs. $\rho$) are shown for beta equilibrated charge neutral neutron star matter.}
\label{symeos}
\end{figure*}
     
    As mentioned above, the nuclear matter EoS is calculated using the isoscalar and the isovector components of M3Y interaction along with density dependence. The energy per nucleon for isospin asymmetric nuclear matter is given by 

\begin{equation}
\epsilon(\rho,X) = [\frac{3\hbar^2k_F^2}{10m}] F(X) + (\frac{\rho J_v C}{2}) (1 - \beta\rho^n)  
\label{seqn1}
\end{equation}
\noindent
where $X$=$\frac{\rho_n - \rho_p} {\rho_n + \rho_p}$ is the isospin asymmetry with $\rho_n$, $\rho_p$ and $\rho$=$\rho_n$+$\rho_p$ being the neutron, proton and nucleon number densities respectively, $k_F$ is Fermi momentum in case of SNM, $F(X)$=$[\frac{(1+X)^{5/3} + (1-X)^{5/3}}{2}]$ and $J_v$ represents the volume integrals of the isoscalar and the isovector parts of the M3Y interaction. The details of the present methodology may be obtained in Ref.\cite{BCS08}. However, for solving the Einstein's equations for stellar structure, we need to consider the total energy density including mass (also called, the mass-energy density) $\varepsilon$ which is related to the $\epsilon$ and baryon number density $\rho$ as $\varepsilon=\rho(\epsilon+m)$ where m ($\sim 938.919$ MeV) is the average of the neutron and proton masses in MeV unit. As the saturation energy per nucleon $\epsilon_0=-15.26$ MeV and the saturation density $\rho_0=0.1533$ fm$^{-3}$ are used in this work, the corresponding total energy density at saturation is $\varepsilon_0$=141.597 MeV/fm$^3$=$2.524\times 10^{14}$ gm/cm$^{3}$. Obviously, as these two parameters $\epsilon_0$ and $\rho_0$ are extracted on the basis of information on finite nuclei, they put the two constraints to the high density behaviour of any nuclear matter EoS. The pressure-density relationship for the present EoS is consistent with the experimental flow data \cite{Da02} confirming its high density behaviour. The values of other important quantities, $L$, $K_{sym}$, $K_{asy}$ and $K_{\tau}$, defined and calculated in Refs.\cite{CBS09,BCS09}, also agree extremely well with those extracted from experiments.

The NSE given by E$_{sym}(\rho)=\epsilon(\rho,1) -\epsilon(\rho,0)$ has a value of $30.71 \pm 0.26$ MeV at the saturation density obtained from this calculation which satisfies one of the constraints on the high density EoS. At higher densities the NSE (see Fig.1(a)) using DDM3Y interaction peaks at $\rho\approx1.95\rho_0$ and becomes negative at $\rho\approx4.7\rho_0$. In Fig.1 the symmetry energy and energy per nucleon (E/A) are plotted as a functions of the baryon density. The equation of state for the $\beta$-stable charge neutral neutron star matter (see Fig.1(b)) is calculated numerically using Eq.(1) with $\beta$-equilibrated proton fraction determined from the NSE. These plots compare the symmetry energy functions and the EoSs determined by the present calculation and several relativistic models (e.g. DD-F, KVOR, DBHF) \cite{Kl06} for the neutron star matter. In Fig.1(a), the NSE calculated by the phenomenological RMF models using density dependent masses and coupling constants (e.g. DD-F, KVOR) goes on increasing with density and never become negative. In the relativistic Dirac-Brueckner-Hartree-Fock (DBHF) approach, the NSE increases more rapidly with density indicating very large proton fraction at higher density. This shows an opposite trend to the NSE function determined from our EoS. The saturation density ($\rho_0$) used in DD-F, KVOR, DBHF and our EoS are 0.1469, 0.1600, 0.1810 and 0.1533 fm$^{-3}$ respectively. So the DBHF uses considerably larger density whereas the $\rho_0$ used by other two models are slightly different from the experimentally extracted value of $0.1533$ fm$^{-3}$. The values of NSE at saturation density calculated by DD-F, KVOR, DBHF and our EoS are 31.6, 32.9, 34.4 and 30.71 MeV respectively. It is clear that DBHF slightly overestimates the value of NSE at $\rho_0$ which is roughly around 30 MeV.

 A negative NSE at high densities implies that the pure neutron matter becomes the most stable state. Consequently, pure neutron matter exists near the core \cite{ab3} of the NSs. Although the present EoS is `stiff' since the SNM incompressibility $K_\infty=274.7\pm7.4$ MeV, but the NSE is `super-soft' because it increases initially with nucleonic density up to about two times the normal nuclear density and then decreases monotonically (hence `soft') and becomes negative (hence `super-soft') at higher densities \cite{BCS08,CBS09}. This is consistent with the recent evidence for a soft NSE at suprasaturation densities \cite{Zh09} and with the fact that the super-soft \cite{We09} nuclear symmetry energy is preferred by the FOPI/GSI experimental data on the $\pi^+/\pi^-$ ratio in relativistic heavy-ion reactions for the stability of NSs.

\begin{figure*}[]
\centerline{\includegraphics[width=3.38in,height=3.38in]{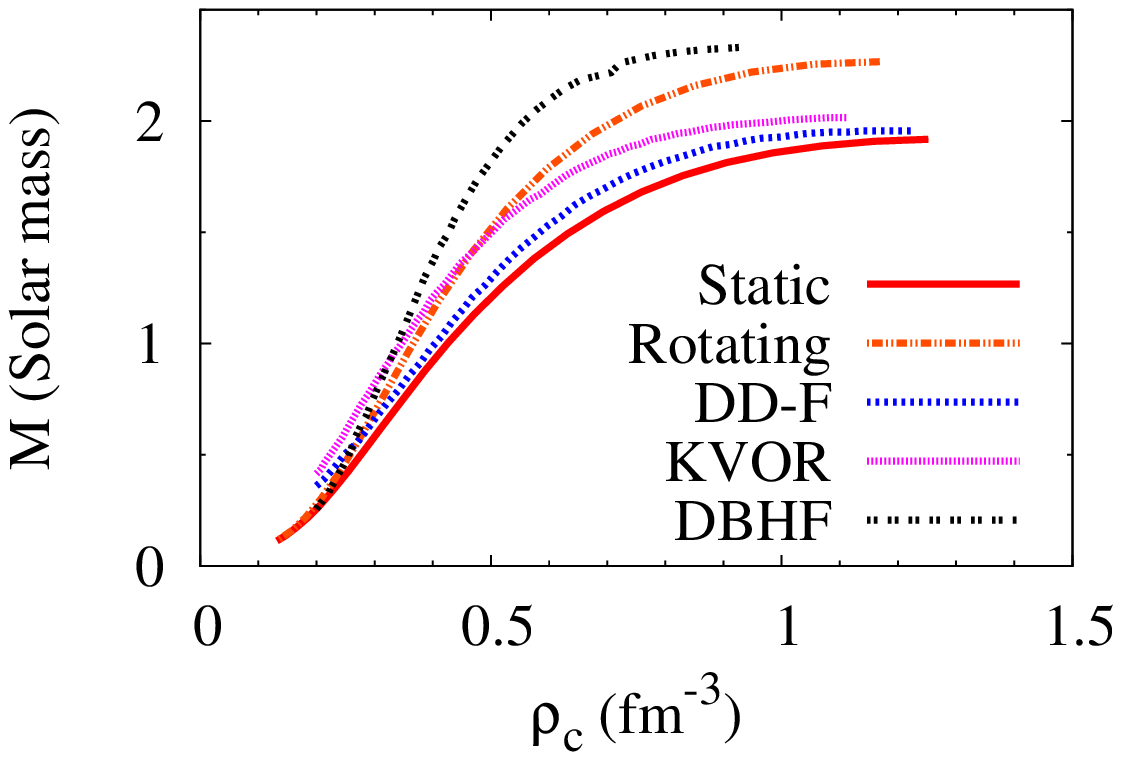}
\includegraphics[width=3.38in,height=3.38in]{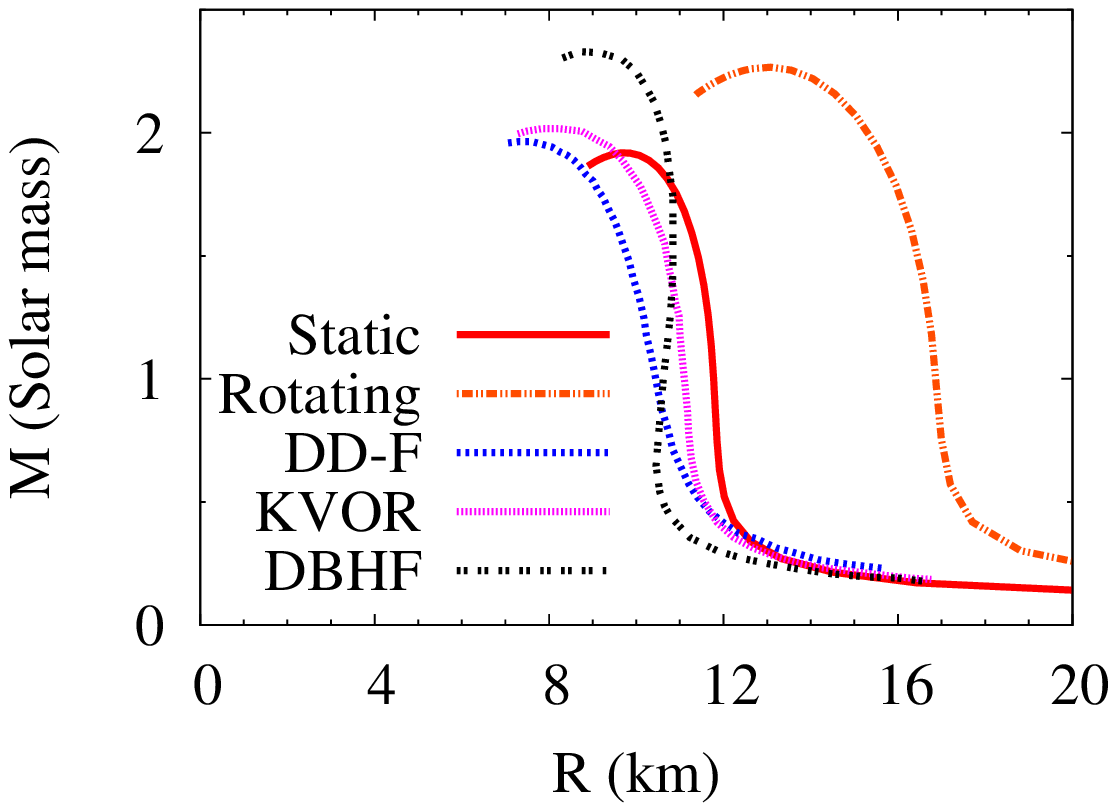}}
\caption{(Color online) (a) (Left) The variation of NS masses (M) is shown with central density ($\rho_c$). The results of this calculation are denoted as `Static' and 'Rotating' for non-rotating and rotating NSs at the Keplerian limit.
(b) (Right) Mass-Radius relationship is given for static and rotating stars at Keplerian frequency using this EoS. The other three plots (DD-F, KVOR, DBHF) present the same relationship for static star only.}
\label{mrhor}
\end{figure*} 
   
The $\beta$ equilibrium proton fraction $x_\beta$ \cite{La91} of a NS consisting of neutrons (n), protons (p) and electrons (e) is completely governed by the density dependent behavior of NSE. Contrary to the relativistic models like DD-F, KVOR, DBHF etc. this work does not support the fast cooling via direct nucleon URCA process as the maximum $x_\beta$ is $4.4\%$ only. Recently it has been concluded theoretically that an acceptable EoS of asymmetric nuclear matter shall not allow the direct URCA process to occur in NSs with masses below 1.5 solar masses \cite{Kl06}. However, the possibility of fast cooling \cite{yakov02,hein09} via direct hyperon URCA or any other processes enhancing neutrino emissivities like $\pi^-$ and $K^-$ condensates may not be completely ruled out. Also a recent experimental observation suggests \cite{cack06} high heat conductivity and enhanced core cooling process indicating the enhanced level of neutrino emission, which may be due to Cooper pairing. Further theoretical studies and sufficient observational data are needed to shed some light on the cooling phenomenon of NS.


    Let us now explore the various properties of static and rotating NSs using the proposed EoS. To study the rotating stars the following metric is used:

\begin{eqnarray}
ds^2=-e^{(\gamma + \rho)}dt^2 + e^{2\alpha}(dr^2+r^2d\theta ^2)\\ \nonumber
 +e^{(\gamma - \rho)}r^2sin^2\theta (d\phi -\omega dt)^2
\end{eqnarray} 
where the gravitational potentials $\gamma, \rho, \alpha$ and $\omega$ are functions of polar coordinates r and $\theta$ only. The Einstein's field equations for the three potentials $\gamma, \rho, \alpha$ have been solved using Green's function technique \cite{KEH,ab1,ab2} and the fourth potential $\omega$ has been determined from other potentials. All the physical quantities may then be determined from these potentials \cite{CST}. The matter inside the NS is approximated as a perfect fluid. Solution of the potentials and hence the calculation of physical quantities can be done numerically. The field equations for rotating stellar structure are solved by using our EoS following the procedure adopted by Komatsu, Eriguchi and Hachisu \cite {KEH,CST}. We choose the `rns' code written by N. Stergioulas \cite {St95} in calculating rotating as well as static NS properties.

In Fig.2(a), we have shown the mass of the stars as a function of central baryon density ($\rho_c$). Our results are plotted for the static and Keplerian limit. This is obvious from the Fig. 2(a) that for the same mass comparatively less central density appears for the rotating stars due to centrifugal action. It may be noted that as angular frequency ($\Omega$) becomes greater, the structure of NS \cite{glen} gets changed not only because of centrifugal flattening, but also because it is taking place against the background of a radially dependent frame dragging frequency. For comparison we have also plotted the results of three other EoSs as mentioned earlier in the text. The maximum mass for the static case is about 1.92~M$_\odot$ with radius $\sim 9.7$~km and for the rotating case it is about 2.27~M$_\odot$ with radius $\sim 13.1$ km. So a mass higher than 1.92~M$_\odot$ would rule out a static star as far as this EoS is concerned. The phenomenological RMF models DD-F and KVOR predict maximum mass around twice solar mass for non-rotating star. The relativistic DBHF model calculates the maximum mass $\sim 2.33$ M$_\odot$ and therefore, the DBHF predicts massive NS even for static case. This is also clear from the Fig.2(b) where the mass-radius relationships for all the above EoSs are shown.

To summarise, we have presented a nuclear EoS at supersaturation densities which satisfies both the constraints from NS and heavy ion collision phenomenology. Our results show that with the stellar configuration, which contain a large fraction of $\beta$ equilibrated NS matter with a thin crust is able to describe highly-massive compact stars, such as the one associated to the millisecond pulsars PSR B1516+02B with a mass M=1.94$^{+0.17}_{-0.19}$ M$_\odot$ (1$\sigma$) \cite{Fe07} and PSR J0751+1807, with a mass M=2.1$\pm$0.2  M$_\odot$ to a 1$\sigma$ confidence level (and 2.1$^{+0.4}_{-0.5}$ M$_\odot$ to a 2$\sigma$ confidence level) \cite{Ni05}. In the case of PSR J1748-2021B, a millisecond pulsar in the Globular Cluster NGC 6440, the measured mass is M=2.74$^{+0.41}_{-0.51}$ M$_\odot$ (2$\sigma$) \cite{fe07}. There are few other EoSs which can explain such a high mass for static case, however, they fail to explain the expected behavior of the NSE. We would like to mention at this stage that a star may not rotate as fast as Keplerian frequency due to r-mode instability \cite{nay06}. There have been suggestions that r-mode may limit the time period to 1.5 ms \cite{ster03}. However, pulsar rotating faster (e.g. PSR J1748−2446ad) than this limit has already been observed \cite{hessels06}. Further observations and a better r-mode modelling may shed more light on this issue.

\noindent

   Modern constraints from the mass and mass-radius-relation measurements require stiff EoS at high densities, whereas flow data from heavy-ion collisions seem to disfavour too stiff behavior of the EoS. The data from massive NSs and pulsars may provide an important cross-check between high-density astrophysics and heavy-ion physics. The variation of pressure with density for the present EoS is consistent with the experimental flow data confirming its high density behaviour. We find that the large values of gravitational masses ($\simeq$2.0~M$_\odot$) for the NSs are possible with the present EoS with the SNM incompressibility $K_\infty=274.7\pm7.4$ MeV, which is rather `stiff' enough at high densities to allow compact stars with large values of gravitational masses $\sim$ 2~M$_\odot$ while the corresponding symmetry energy is `super-soft' as preferred by FOPI/GSI experimental data. Thus the DDM3Y effective interaction which is found to provide unified description of elastic and inelastic scattering, various radioactivites and nuclear matter properties, also provides excellent description of $\beta$ equilibrated NS matter to allow the recent observations of the massive compact stars. 

    The research work of P. Roy Chowdhury is sponsored by the UGC (No.F.4-2/2006(BSR)/13-224/2008(BSR)) under Dr. D.S. Kothari Postdoctoral Fellowship Scheme. The work of A. Bhattacharyya is partially supported by UGC (UPE $\&$ DRS) and CSIR.


\end{document}